\documentclass[prd,twocolumn,showpacs,letterpaper,nofootinbib]{revtex4}
\usepackage{amssymb,amsmath,amsfonts}
\usepackage{color}
\usepackage{graphicx}

\newcommand{\bn}{\hat{\mathbf{n}}}
\newcommand{\br}{\mathbf{r}}

\newcommand{\bfH}{{\mathbf{H}}}
\newcommand{\bfk}{{\mathbf{k}}}

\newcommand{\beq}{\begin{equation}}
\newcommand{\eeq}{\end{equation}}
\newcommand{\bea}{\begin{eqnarray}}
\newcommand{\eea}{\end{eqnarray}}
\newcommand{\ba}{\begin{array}}
\newcommand{\ea}{\end{array}}

\newcommand{\PRL}{Phys. Rev. Lett.}
\newcommand{\PRD}{Phys. Rev. D}

\newlength{\sizeonefig}
\newlength{\sizetwofig}
\begin{document}

\title{Cosmic Shear from Scalar-Induced Gravitational Waves}

\author{Devdeep Sarkar$^1$, Paolo Serra$^1$,  Asantha  Cooray$^{1}$, Kiyotomo  Ichiki$^2$,  Daniel  Baumann$^3$}  

\affiliation{
$^1$Center for  Cosmology, Department  of Physics and  Astronomy, 4129
Frederick Reines  Hall, University  of California, Irvine,  CA 92697\\
$^2$Research Center for the Early Universe, University of Tokyo, 7-3-1
Hongo,  Bunkyo-ku, Tokyo  113-0033, Japan\\  
$^3$Department of Physics, Princeton University, Princeton, NJ 08544}

\date{\today}

\begin{abstract}
Weak  gravitational lensing by  foreground density perturbations
generates  a gradient  mode  in  the shear  of  background images.
In contrast, cosmological tensor perturbations induce a  non-zero  
curl mode associated with image rotations.
In this note, we study the lensing signatures of both primordial gravitational waves from inflation and second-order gravitational waves generated from the observed spectrum of primordial density fluctuations.
We derive the curl mode for  galaxy  lensing surveys  at
redshifts of 1  to 3 and for lensing of  the cosmic microwave background (CMB)
at a  redshift of  1100.  
We find that the curl mode angular power spectrum associated
with secondary tensor modes for  galaxy lensing surveys  dominates 
over the corresponding signal generated by primary gravitational
waves from inflation.
However, both tensor contributions to the shear curl mode spectrum are below the projected noise levels of upcoming  galaxy and  CMB lensing surveys and therefore are unlikely to be detectable.

\end{abstract}

\maketitle

\section{Introduction}

The weak lensing of background sources such as galaxies at redshifts
of 1 to 3 and cosmic microwave background
(CMB) fluctuations at a redshift of 1100 by foreground density perturbations is now
well understood \cite{lensing,Cha,cosmicshearrefs}.
In  addition  to the lensing by density  perturbations,  metric tensor  perturbations
associated  with  gravitational  waves  also  lens  background  images
\cite{Jaffe,stebbins}.   While  the lensing  by
gravitational   waves   was  first   considered   to  be   negligible
\cite{Jaffe}, the advent of  high precision weak lensing surveys
(both from the ground and from space)  as well as the potential availability of
high resolution  and high sensitivity CMB  anisotropy and polarization
maps  has   renewed  interest  in the  lensing  by  gravitational  waves
\cite{DRS03,LiCoo}.

An  important  source  of  cosmological  gravitational  waves  are quantum fluctuations during the inflationary era.
   The weak lensing  of background
galaxy images  \cite{DRS03} and CMB
anisotropies \cite{LiCoo}  by these primordial tensor modes  has
previously been discussed in  the literature.  Even for the maximal inflationary gravitational wave amplitude consistent with current observations (corresponding to a tensor-to-scalar ratio, $r \lesssim 0.4$),  the weak lensing
effect on galaxy images is below the noise level even for a next-generation all-sky lensing
survey and is therefore unlikely to be detectable \cite{DRS03}. For lensing of CMB
anisotropies and polarization, the modifications imposed by foreground
gravitational waves  with a tensor-to-scalar ratio below  0.4 is again
smaller  than  the cosmic  variance  for  all-sky  CMB anisotropy  and
polarization measurements \cite{LiCoo}.

While previous studies have concentrated on the lensing  by first-order primordial
gravitational waves, a secondary spectrum of gravitational waves is generated at by second-order by the observed primordial density
fluctuations \cite{GW2}.  These tensors  produce a B-mode  spectrum in
the CMB  polarization \cite{KamKosSte97} with  an equivalent amplitude
that  is   about  $10^{-6}$  in  the  (first-order) tensor-to-scalar  ratio,  after
accounting for late-time reionization contribution to CMB polarization
\cite{Moletal}.   In the presence of residual polarized  foregrounds and the confusion produced by weak lensing  of CMB anisotropies by foreground density perturbations
\cite{KesCooKam02}, such a signal is in practice unobservable. 
Furthermore, the present  amplitude of secondary scalar-induced gravitational  waves is below the projected sensitivity for future
experiments  like   the  Big  Bang  Observer   (BBO)  at
the wavelengths corresponding to  space-based direct detection experiments \cite{Ananda}.  However, on larger scales secondary gravitational waves are continuously sourced by a non-linear scalar source term. As a consequence secondary gravitational
waves have a non-trivial transfer function and the late-time
  spectrum is enhanced on  cosmological length scales relative to the small scales accessible to direct detection experiments \cite{Bauetal}. 
  In particular, on comoving scales of order the horizon size at matter-radiation equality ($\sim k^{-1}_{\rm eq}$) second-order gravitational wave does not redshift and their amplitude stays constant.  This is in contrast to (first-order) primordial gravitational waves that redshift on all scales.
This effect leads to a peak of the secondary gravitational wave spectrum on large scales (around $k_{\rm eq}$) which could 
potentially be probed with weak  lensing of galaxies at redshifts of 1
to 3 (see Figures 1 and 3 in Ref.~\cite{Bauetal}). 

The identification of lensing by gravitational waves is aided by
the fact that the lensing deformation associated with tensors leads to a curl mode in 
cosmic  shear
\cite{stebbins,CooKamCal05}.  Foreground  density perturbations do not
generate a  curl mode in cosmic  shear, except at second-order  and at
small  angular  scales  due  to  effects such  as  lens-lens  coupling
\cite{CooHu02}. The  situation is analogous  to  the curl (B)  and the  gradient
(E) modes  of CMB polarization, where only gravitational waves source
the curl or B-mode \cite{KamKosSte97}.

The paper is organized as follows. In Section~II we review theoretical aspects of lensing by foreground gravitational waves.  Computing the lensing signals requires input power spectra and transfer functions for 
both primordial tensors and the secondary tensors sourced by primordial density perturbations. We provide these results in Section~III.
In Section~IV we present our results on the shear curl mode
angular power spectrum. We conclude in Section~V. 

When presenting numerical calculations,
we will assume  a flat-$\Lambda$CDM cosmology with $\Omega_m=0.3$ and $h=0.7$.

\section{Weak Lensing by Gravity Waves}

In weak gravitational lensing, density perturbations only
lead  to  image  distortions  involving amplification  (or  convergence)
$\kappa$.  However, the lensing  by gravitational  waves leads  to both
convergence and  rotation $\omega$, involving  the anti-symmetric part
of the weak lensing deformation matrix \cite{CooHu02}
\begin{equation}
\label{A}
{\bf A} =
\left(
\begin{array}{cc}
1-\kappa-\gamma_1 & -\gamma_2-\omega\\
-\gamma_2+\omega & 1-\kappa+\gamma_1
\end{array}
\right) \, ,
\end{equation}
where all  components are functions of  the position on  the sky $\hat {\bf
n}$ and $\gamma_i$ are 
two shear components \cite{cosmicshearrefs}.
$({\bf A})_{ij} \equiv A_{ij}$ maps between the source plane (S) and the image plane (I) such
that $\delta  x_i^S=A_{ij}\delta x_j^I$. 

For lensing by foreground gravitational waves, the geodesic
equation is \cite{DRS03}
\begin{eqnarray}
\ddot{\bf r} &=& \frac{1}{2}\left(\dot{{\bf r}} \cdot \dot{\bf H} \cdot \dot{\bf r}\right) \dot{\bf r} \nonumber \\
 & & -\left({\bf 1}+{\bf H}\right)^{-1} \cdot
\left[ \dot{\bf r} \cdot \frac{d}{{d} \eta}{\bf H} -
\frac{1}{2} \nabla_{\hspace{-0.08cm} H} \left(\dot{\bf r} \cdot {\bf H} \cdot \dot{\bf r}\right)\right] \, ,
\label{eq:rdotdot}
\end{eqnarray}
where, to simplify notation,  the explicit
dependence    on    $\eta$   in    each  of  these  terms has been suppressed.    
Here, over-dots  represent derivatives with
respect  to  conformal  time  and $({\bf  H})_{ij} \equiv h_{ij}$  is  the  transverse
($\nabla\cdot\bfH=0$)  and   traceless  (Tr\,$\bfH=0$)  tensor  metric
perturbation representing gravitational waves.
The operator $\nabla_{\hspace{-0.08cm} H} $ denotes
the gradient applied only to the metric  perturbation ${\bf H}$;
when not subscripted  with $H$, the gradient should  be interpreted as
applying to all  terms, including the  line-of-sight
directional vector  $\bn$. 
The solution to the above equation, ${\bf r}(\bn, \eta)$, is discussed
in Refs.~\cite{DRS03,LiCoo}.

Using  the transverse  displacement associated  with a perturbed photon
trajectory, the  angular deflection projected onto the sky is
$\vec{\Delta} = [{\bf r} - ({\hat{\bf n}} \cdot {\bf r})\hat{\bf
n}]/(\eta_0-\eta)$.
This 
can  be
related to 
the convergence $\kappa$ and the rotation $\omega$ in the
weak lensing
deformation matrix \cite{stebbins} 
\bea \kappa\equiv-\frac{\Delta^a{}_{:a}}{2},\ \quad
{\rm  and} \quad  \ \omega\equiv\frac{(\Delta_a\epsilon^{ab})_{:b}}{2}
\, , \eea  
where the colons denote covariant derivatives with respect to the perturbed FRW metric \cite{stebbins}.

A simple  argument explains why gravitational waves $h_{ij}$ lead to an image rotation.
If we take the line-of-sight to be in the $\hat{\bf z}$-direction, then  $\omega \propto \epsilon_{kl}  \partial_k h_{zl}$. If
the gravitational wave propagates in the $\hat{\bf y}$-direction, then
$\omega  \propto  \partial_y h_{zx}$  and  the  deflection  is in  the
$\hat{\bf x}$-direction with $\delta \theta_x \propto h_{zx}$.

Integrating over all  deflections along
the line-of-sight to a background image at $\eta_S$, we can write the
rotational component as \cite{stebbins}
\begin{eqnarray}
\omega(\bn)&\equiv&-\frac{1}{2}\bn\cdot(\nabla\times\br(\bn,\eta_S)) \nonumber \\
&=&\frac{1}{2}\int_{\eta_S}^{\eta_0} {\rm d} \eta' \left[\bn\cdot(\nabla
\times {\bf H}) \cdot \bn\right]  \, . 
\label{BornOmega}
\end{eqnarray}
Assuming  isotropy, the  three-dimensional spatial  power  spectrum of
initial  metric fluctuations  related  to a  stochastic background  of
gravitational waves is
\begin{equation}
\langle {h}_{(i)}(\bfk) {h}_{(j)}^*(\bfk') \rangle
=(2\pi)^3P_t(k)\,\delta_{ij}\,\delta^{(3)}(\bfk-\bfk')\ ,
\end{equation}
where the two linear-polarization states of the gravitational wave are
denoted by  $(i),(j)= \times, +$. 
Taking  the spherical-harmonic  moments of  Eq.~(\ref{BornOmega}) and using $(\nabla  \times {\bf
H})_{il}  =  \epsilon_{ijk}  \partial_j  h_{kl}$,  the  angular  power
spectrum of the rotational component becomes \cite{DRS03,LiCoo}
\begin{eqnarray}
C_l^{\omega \omega} &=& {1\over2l+1}\sum_{m=-l}^l
\left\langle|{\omega}_{lm}|^2\right\rangle \nonumber \\
&=&{2 \over \pi}\int k^2\,{\rm d} k\,P_t(k)\,\left|T_l^\omega(k,\eta_S)\right|^2\, ,
\end{eqnarray}
where
\begin{eqnarray}
\label{LensingTransfer}
\hspace{-1.cm}T_l^\omega(k,\eta_S)&=& \nonumber  \\
&& \hspace{-2.cm} = \sqrt{{(l+2)!\over(l-2)!}}\,
  \int_{\eta_S}^{\eta_0} k \, {\rm d}\eta'\,T_t(k,\eta')\,
  \left. {\frac{j_l(x)}{x^2}}\right|_{x=k(\eta_0-\eta')} \hspace{-1cm}.
\end{eqnarray}
Here  $T_t(k,\eta)$  represents the  transfer  function of  tensor
perturbations.   

\begin{figure}[!t]
\includegraphics[scale=0.50,angle=0]{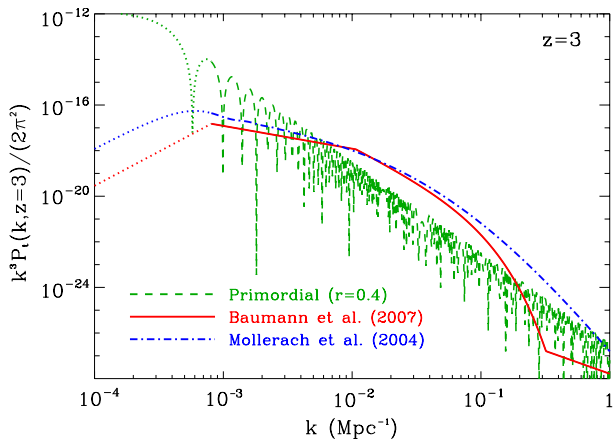}
\caption{The power spectrum  of primordial (dashed  line) and secondary
gravitational waves at $z=3$.  For the secondary spectrum, we show results
from two  calculations in the  literature: the solid line is from  Baumann et
al.   \cite{Bauetal}   and   the dot-dashed   line   from   Mollerach   et
al. \cite{Moletal}.  The dotted part of the spectra corresponds to superhorizon scales at $z=3$.}
\label{pk}
\end{figure}

\section{Secondary Tensor Spectrum}

The  derivation of  the cosmic shear  curl modes  from a  spectrum  of tensor
fluctuations  has so  far made  no reference  to the form of the underlying power
spectrum $P_t(k)$ and the transfer function $T_t(k,\eta)$. 
The previous results are therefore applicable to different sources  for cosmological  gravitational waves.
In  standard inflationary  models, the  primordial  tensor fluctuation
spectrum is predicted to be 
\begin{equation}
P_t(k) = A_t k^{n_t-3} \ .
\end{equation}
Inflationary  models generally  predict that  $n_t \sim  0$  while the
ratio of tensor-to-scalar  amplitudes, $r=A_t/A_s$, is now constrained
to be  $\lesssim 0.4$  \cite{WMAP3}.  We will  use this upper
limit   when   calculating   the   inflationary   gravitational   wave
contribution to shear curl modes.  In addition to the primordial power
spectrum,  we also  require the  transfer  function $T_t(k,\eta)$.
This  is obtained as a  solution to  the  wave equation for primordial gravitational waves \bea
\ddot{{\bf H}}  - \nabla^2 {\bf H}  + 2 \frac{\dot{a}}{a}  {\bf H}= 16
\pi G a^2  {\bf P} \, ,\eea where  ${\bf P}$ is the tensor  part of the
anisotropic stress,  say due to neutrinos  \cite{Prit}. The term on the
right  hand  side  acts  as  a  damping  term  for  the  evolution  of
gravitational waves and is important  for modes that enter the horizon
before matter-radiation equality, with  a smaller correction for modes
entering the  horizon after matter-radiation  equality.  Ignoring this
small  correction,  we  take  the  transfer  function  for  the primordial
gravitational wave spectrum as $T_t(k,\eta)=3j_1(k\eta)/(k\eta)$.

We now consider the spectrum and transfer function for
cosmological
gravitational waves which are created at second-order by the observed primordial density perturbations \cite{GW2}.  We  make use of
two calculations  in the literature for the spectrum  of 
secondary  gravitational  waves.   Using  results  from  Mollerach et
al.~\cite{Moletal}, the secondary tensor power spectrum is given by
\begin{equation}
P_t^{\rm Mol}(k) = \frac{12\pi^2}{25}C \Delta_{\mathcal R}^4 (k_0) \frac{1}{k^3} \frac{k_*}{k} W(k/k_*) \left(\frac{k_*}{k_0}\right)^{2(n_s-1)} ,
\end{equation}
with $k_*  = \Omega_{m}h^2  \, \text{Mpc$^{-1}$}$ and  the coefficient
$C(n_s)=0.062$ when  $n_s=1$.  The function  $W(x)$ is well  fitted by
$W(x) =  (1+7x+5x^2)^{-3}$.  The normalization of the scalar spectrum is  taken  to  be  $\Delta_{\mathcal  R}^2  (k_0  =  0.002  \,
\text{Mpc$^{-1}$}) =  2.4 \times 10^{-9}$ \cite{WMAP3} and
to simplify  the calculation  we assume a  spectral index  for density
perturbations with  $n_s \sim 1$.  Our results  and conclusions are insensitive to assuming $n_s  \sim 0.96$, more consistent  with recent
WMAP results \cite{WMAP3}.  The transfer function associated with this
secondary gravitational wave spectrum is
\begin{equation}
T_t^{\rm Mol}(k,\eta) = \left( 1 - \frac{3j_1(k\eta)}{k\eta}\right)g_{\infty}^2 , 
\end{equation}
where $g_{\infty}$  is the growth-suppression  factor in the  limit $z
\rightarrow \infty$ \cite{CPT}.

\begin{figure*}[!t]
\includegraphics[scale=0.50,angle=0]{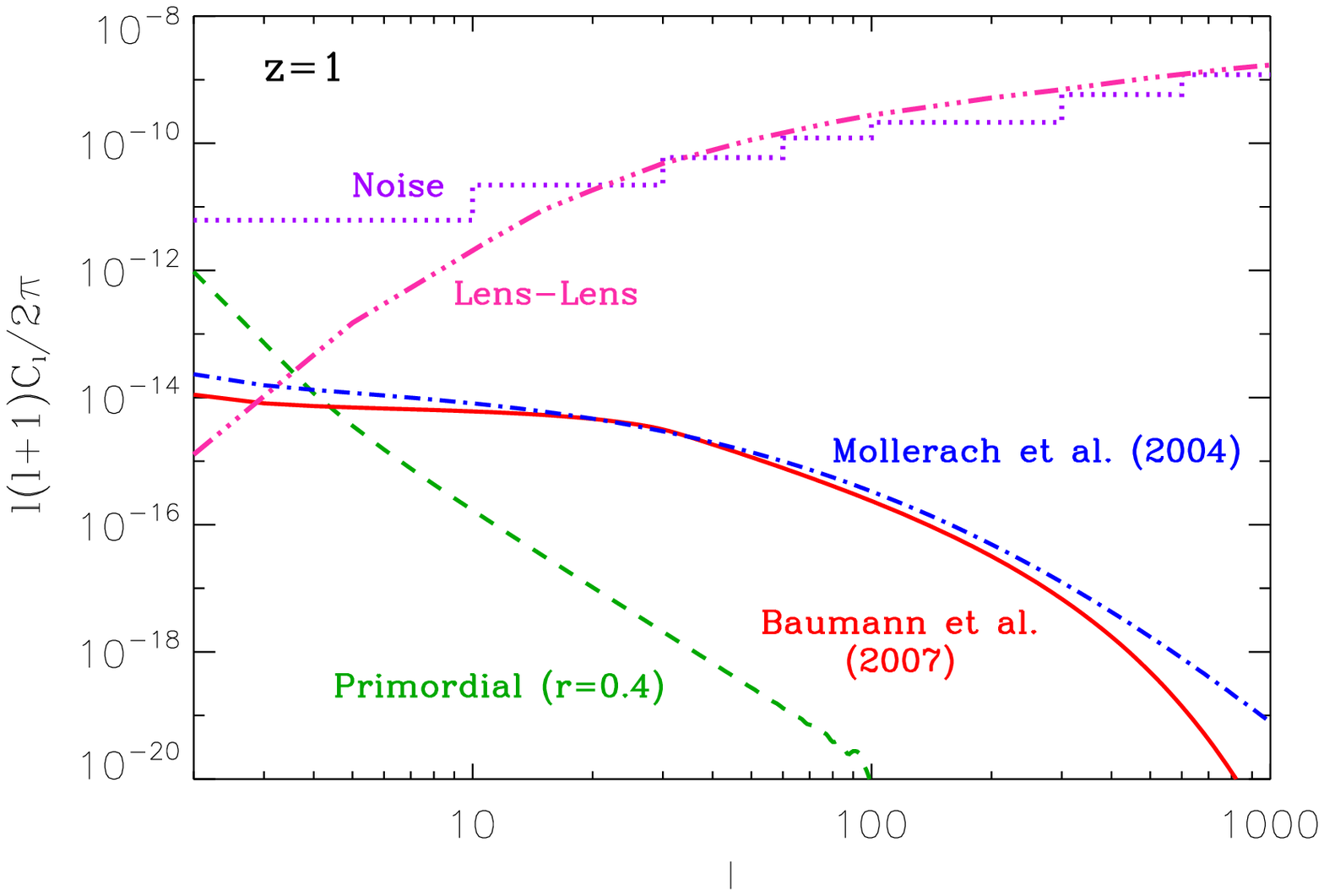}
\includegraphics[scale=0.50,angle=0]{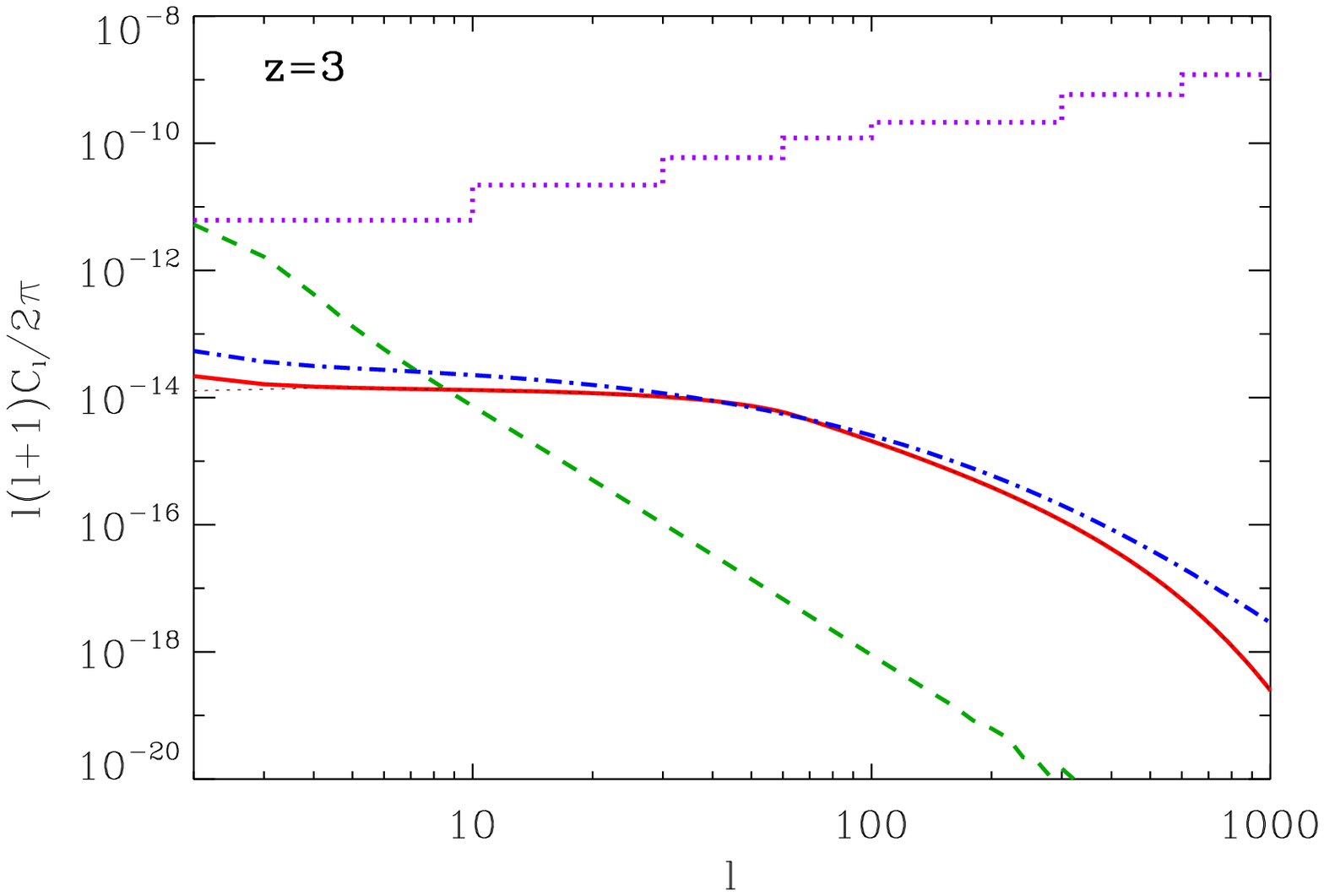}
\caption{The angular power spectra of the cosmic shear curl mode at $z=1$
(left panel)  and at $z=3$ (right panel). The dotted line  is the noise
associated with  a measurement of the shear curl mode power spectrum (see text
for details). While the curl  mode power spectrum is below the noise, the
secondary  gravitational  waves  produce  a  larger  signal  than  the
primordial tensor modes when $r < 0.4$. For reference, the double dot-dashed line on the left-panel shows the angular power spectrum of secondary
shear curl modes generated by the coupling of two lenses (lens-lens coupling) along the line-of-sight to
background sources at $z=1$ \cite{CooHu02}.}
\label{clgalaxy}
\end{figure*}

The Mollerach  et al.  \cite{Moletal} calculation of  secondary tensor
fluctuations was recently extended by Baumann et al.~\cite{Bauetal} by
accounting  for  the evolution  over   all  wavenumbers  during  both
radiation  and  matter  domination.  The  analytical  result  for  the
scalar-induced gravitational wave power spectrum is \cite{Bauetal}
\begin{eqnarray}
P_t^{\rm Bau}(k) &=& 2\pi^2 \left(\frac{4}{9}\right)^2 \Delta_{\mathcal R}^4(k_0) k^{-3} \left(\frac{k}{k_0}\right)^{2(n_s-1)} \nonumber \\
& & \times 
\left\{
\begin{array}{rl}
  \frac{k_{\rm eq}}{k} & \text{if } k < k_{\rm eq} \\
  1 &  \text{if } k > k_{\rm eq} 
\end{array} \right. 
\end{eqnarray}
while the corresponding transfer function is
\begin{equation}
T_t^{\rm Bau}(k,\eta) =  \left\{
\begin{array}{ll}
        1                                 & \text{if } k < k_{\rm eq} \\
  \left(\frac{k}{k_{\rm eq}}\right)^{-\gamma(k)} & \text{if } k_{\rm eq} < k < k_c(\eta)\\
   \frac{a_{\rm eq}}{a(\eta)}\frac{k_{\rm eq}}{k} & \text{if } k > k_c(\eta)
\end{array} \right.
\end{equation}
where
\begin{equation}
k_c(\eta) = \left[ \frac{a(\eta)}{a_{\rm eq}} \right]^{1/(\gamma(k)-1)}k_{\rm eq} \, .
\end{equation}
Here, $k_{\rm eq}  = 0.073\, \Omega_m h^2$  Mpc$^{-1}$ corresponds to the comoving horizon scale at matter-radiation equality. 
$\gamma(k) $ is a weakly $k$-dependent function which we fit by
 comparison to
numerical calculations of the tensor power spectrum in Baumann et al.~\cite{Bauetal}.  In practice, we use a smooth interpolation between $k_{\rm eq}$ and $k_c(\eta)$.  For low $z$ we find $\gamma(k_{\rm eq})=1.5$ and $\gamma(k_c)=3$.  The  analytical result presented here was found to be in agreement with full numerical result
at the  10\% level and is adequate for the purposes of this calculation. 

\begin{figure}[t]
\includegraphics[scale=0.50,angle=0]{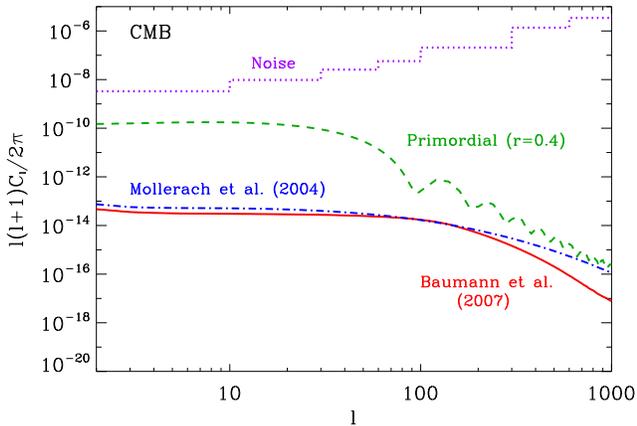}
\caption{The angular power spectrum  of the curl mode at $z=1100$ for lensing  of   CMB  anisotropies   by  foreground
gravitational waves. The  dotted line shows the expected  noise from a
cosmic-variance  limited  reconstruction of  the  curl mode  following
Cooray et al. \cite{CooKamCal05}  using E- and B-mode CMB polarization
maps. For  CMB lensing, the  primordial tensor  modes dominate
when $r \gtrsim 10^{-6}$.  }
\label{clcmb}
\end{figure}

\section{Results and Discussion}

In Figure 1, we  give a comparison between the  primordial gravitational wave
spectrum  with  a tensor-to-scalar  ratio  of  0.4  and the  secondary
gravitational  wave spectrum  at   $z=3$.  We  show  results from  both
Mollerach et al. \cite{Moletal} and Baumann et al.~\cite{Bauetal} for the second-order tensor spectrum.  They agree at the 
percent level when $k < k_{\rm eq} (=0.0107\, {\rm Mpc}^{-1})$ and at high redshifts.
 For smaller scales, $k > k_{\rm eq}$, and at  low  redshifts,  due  to  differences  in  the
treatment of the  evolution of the tensor modes  captured in the
transfer function,  the two  calculations  predict spectra that  differ by  more than  a few  percent.  
Baumann et al.~\cite{Bauetal} captures the correct transfer function for small scale gravitational waves.

In  Figure 2, we show  the weak lensing curl mode  angular power  spectrum for
sources out to $z=1$ and at  $z=3$. 
The dotted line in  Fig. 2 shows the
expected binned  noise from an  all-sky experiment similar to  the one
discussed in  Dodelson et al.  \cite{DRS03}.  For weak  lensing shear,
the binned noise is
\begin{equation}
\Delta C_l = \sqrt{\frac{2}{(2l+1)\Delta_l f_{\rm sky}}}\frac{\left<\gamma^2\right>}{N_{\rm gal}} \, ,
\end{equation}
and we take $\langle \gamma \rangle$, the intrinsic ellipticity, to be
0.3 and $N_{\rm  gal}=1.5 \times 10^{10}$ or roughly  100 galaxies per
square-arcminute.  The plotted noise  power spectrum in Fig.~2 assumes
varying bin  sizes, $\Delta_l$, but  as shown  there, even with  wide bins  in the
multipole space, the detection of secondary tensor modes with the curl
mode of  cosmic shear remains  challenging. 

Even if we there were a technique to improve on the measurement noise of lensing surveys, 
the signal from secondary tensors will be heavily confused with
another signal in the shear curl mode associated with the coupling of two lenses
along the line-of-sight (lens-lens coupling; \cite{CooHu02}). We show the resulting  angular
power spectrum out to $z=1$ in the left panel of Fig.~2 with a double dot-dashed line.
This signal  peaks at small angular scales as it is generated by non-linear density
perturbations.
At multipoles of 10 to 100 where the secondary tensor signal is
interesting the lens-lens coupling corrections to the rms of the curl mode is larger by
a factor of 10 to 100. 
%In a high signal-to-noise ratio map of the curl mode, the
%lens-lens coupling contribution can be established by cross-correlating such a map 
%with a map of the foreground large-scale structure. 

While the signal  is below 
the measurement noise and is confused with the lens-lens coupling signal in the curl modes of shear, 
the secondary tensor modes  produce a larger curl mode at $\ell
\sim  100$  than the  primordial  tensor  modes  from inflation  with
$r \lesssim 0.4$.
Thus, we  find that at  large angular
scales the curl  modes of cosmic shear will  be dominated by secondary
gravitational waves  and not the primordial  signal from inflation. This
is consistent with results in  Baumann et al. \cite{Bauetal} which show
that  at cosmologically  interesting  wavenumbers with  $k\sim10^{-3}$
Mpc$^{-1}$  to 0.1  Mpc$^{-1}$,  the secondary  spectrum dominates over  the
primordial spectrum at low redshifts.  While such modes are not probed
by a  direct detection experiment such  as the BBO, such  modes are in
the range that is in principle detectable with  cosmic shear.  Unfortunately, the amplitude
is below what can  be achieved with galaxy lensing surveys, even
considering optimistic galaxy statistics  and shear noise.

Finally, in Fig.~\ref{clcmb}, we plot  $C_l$ of cosmic shear for $z=1100$ related to
lensing of  CMB anisotropies  by foreground tensors  \cite{LiCoo}. The
noise plotted here comes from a cosmic variance-limited reconstruction
of  the  shear  curl  mode  with  a  combination  of  E-  and  B-mode
polarization  maps   \cite{CooKamCal05,spin2}.   For  both  primordial
gravitational waves and the secondary gravitational waves, a detection
is unlikely  to be  achieved. In  the case of  the CMB,  unlike galaxy
lensing  at low redshifts,  the primordial  tensors dominate  the curl
mode  (for $r \gtrsim 10^{-6}$) since one is  probing out  to a  high redshift  where primordial
modes are not significantly damped due to subsequent evolution.

\section{Conclusion}

At second-order in perturbation theory the measured spectrum of primordial density fluctuations generates a secondary gravitational wave signal.
In this paper, we computed the weak lensing signatures of these secondary tensor modes.
 We considered the use of the
cosmic shear curl mode, or  analogously the rotational component, as a
diagnostic of these tensor modes since density perturbations at first-order do not generate a curl mode.  
We presented results both  for galaxy  lensing
surveys  at  redshifts of  1  to 3  and  lensing  of cosmic  microwave
background (CMB) fluctuations at a redshift  of 1100.  
At low redshifts,  the signal associated
with secondary  tensor modes is larger than the  shear curl
mode from primary gravitational waves generated by inflation
with a tensor-to-scalar ratio less than 0.4.
However, we find that the expected shear curl mode
spectrum from both primordial and secondary gravitational waves  is very small and unlikely to be detectable with upcoming galaxy and  CMB lensing surveys.

\smallskip

\begin{center}

{\bf Acknowledgments\\}

\end{center}

This work was supported by NSF CAREER AST-0645427 at UC Irvine.
KI is supported by a Grant-in-Aid for the Japan Society 
for the Promotion of Science. DB thanks the Perimeter Institute for hospitality during completion of this work.


\vfill


\begin{thebibliography}{99}

\bibitem{lensing}  See, e.g., U.~Seljak and M.~Zaldarriaga, \PRL\ {\bf
  82}, 2636 (1999); \PRD\ {\bf 60},
  043504 (1999); M. Zaldarriaga and
  U. Seljak, \PRD\ {\bf 59}, 123507 (1999); W.~Hu, \PRD\ {\bf 64}, 083005 (2001); W.~Hu, \PRD\ {\bf 62}, 043007  (2000);
  M.~Zaldarriaga and U.~Seljak,
  %``Gravitational Lensing Effect on Cosmic Microwave Background Polarization,''
  Phys.\ Rev.\ D {\bf 58}, 023003 (1998).
  %%CITATION = ASTRO-PH 9803150;%%
%  W.~Hu and A.~Cooray,
  %``Gravitational time delay effects on cosmic microwave background
  %anisotropies,''
 % Phys.\ Rev.\ D {\bf 63}, 023504 (2001).
  %%CITATION = ASTRO-PH 0008001;%%  

%\cite{Lewis:2006fu}
\bibitem{Cha}
A.~Lewis and A.~Challinor,
  %``Weak Gravitational Lensing of the CMB,''
  arXiv:astro-ph/0601594.
  %%CITATION = ASTRO-PH 0601594;%%


\bibitem{cosmicshearrefs} R. D. Blandford et al.,
     Mon.\ Not.\ Roy.\ Astron.\ Soc.\ {\bf 251}, 600 (1991);
     J. Miralda-Escud\'e, Astrophys.\ J.\ {\bf 380}, 1 (1991);
     N.~Kaiser, Astrophys.\ J.\  {\bf 388}, 277 (1992);
     M. Bartelmann and P. Schneider, Astron.\ Astrophys.\ {\bf                                                                                                          
     259}, 413 (1992).
%For reviews, see,
%M.~Bartelmann and P.~Schneider, Phys. Rept. {\bf 340}, 291 (2001);
%P.~Schneider  Gravitational Lensing: Strong, Weak \& Micro,
%Lecture Notes of the 33rd Saas-Fee Advanced Course, (Berlin:
%Springer-Verlag).

\bibitem{Jaffe}
  N.~Kaiser and A.~H.~Jaffe,
  %``Bending of light by gravity waves,''
  Astrophys.\ J.\  {\bf 484}, 545 (1997).
  %%CITATION = ASTRO-PH 9609043;%%

\bibitem{stebbins} A. Stebbins, arXiv:astro-ph/9609149.

\bibitem {DRS03}
 S.~Dodelson, E.~Rozo and A.~Stebbins,
  %``Primordial gravity waves and weak lensing,''
  Phys.\ Rev.\ Lett.\  {\bf 91}, 021301 (2003).
  %%CITATION = ASTRO-PH 0301177;%%

\bibitem{LiCoo}
  C.~Li and A.~Cooray,
  %``Weak lensing of the cosmic microwave background by foreground
  %gravitational waves,''
  Phys.\ Rev.\  D {\bf 74}, 023521 (2006).
  %%CITATION = PHRVA,D74,023521;%%

\bibitem{GW2}
  S.~Matarrese, O.~Pantano and D.~Saez,
% ``A General relativistic approach to the nonlinear evolution of collisionless matter,''
  Phys.\ Rev.\ D {\bf 47}, 1311 (1993);
  S.~Matarrese, O.~Pantano and D.~Saez,
%``General relativistic dynamics of irrotational dust: Cosmological implications,''
  Phys.\ Rev.\ Lett.\  {\bf 72}, 320 (1994);
  S.~Matarrese, S.~Mollerach and M.~Bruni,
 % ``Second-order perturbations of the Einstein-de Sitter universe,''
  Phys.\ Rev.\ D {\bf 58}, 043504 (1998);
  H.~Noh and J.~c.~Hwang,
%  ``Second-order perturbations of the Friedmann world model,''
  Phys.\ Rev.\ D {\bf 69}, 104011 (2004);
  C.~Carbone and S.~Matarrese,
%  ``A unified treatment of cosmological perturbations from super-horizon to small scales,''
  Phys.\ Rev.\ D {\bf 71}, 043508 (2005);
  K.~Nakamura,
%  ``Second-order gauge invariant cosmological perturbation theory: Einstein
%  equations in terms of gauge invariant variables,''
  Prog.\ Theor.\ Phys.\  {\bf 117}, 17 (2007).
    
\bibitem{KamKosSte97} M.~Kamionkowski, A.~Kosowsky, and
     A.~Stebbins, \PRL\ {\bf 78}, 2058 (1997);
        U.~Seljak and M.~Zaldarriaga, \PRL\ {\bf 78}, 2054 (1997).


\bibitem{Moletal}
  S.~Mollerach, D.~Harari and S.~Matarrese,
  %``CMB polarization from secondary vector and tensor modes,''
  Phys.\ Rev.\  D {\bf 69}, 063002 (2004).
  %%CITATION = PHRVA,D69,063002;%%

%\bibitem{Bock}
 % J.~Bock {\it et al.},
 % ``Task Force on Cosmic Microwave Background Research,''
 % arXiv:astro-ph/0604101.

%\bibitem{Amblard}
 % A.~Amblard, A.~Cooray and M.~Kaplinghat,
  %``Search for gravitational waves in the CMB after WMAP3: Foreground
  %confusion and the optimal frequency coverage for foreground  minimization,''
 % Phys.\ Rev.\  D {\bf 75}, 083508 (2007).
  %%CITATION = PHRVA,D75,083508;%%


\bibitem{KesCooKam02} M.~Kesden, A.~Cooray, and M.~Kamionkowski,
     \PRL\ {\bf 89}, 011304 (2002);
     L.~Knox and Y.-S.~Song, \PRL\ {\bf 89}, 011303 (2002); U. Seljak and C. Hirata,
     Phys. Rev. D {\bf 69}, 043005 (2004).


%\bibitem{Smith1}
 % T.~L.~Smith, M.~Kamionkowski and A.~Cooray,
  %``Direct detection of the inflationary gravitational wave background,''
 % Phys.\ Rev.\  D {\bf 73}, 023504 (2006).
  %%CITATION = PHRVA,D73,023504;%%

%\cite{Ananda:2006af}
\bibitem{Ananda}
  K.~N.~Ananda, C.~Clarkson and D.~Wands,
  %``The cosmological gravitational wave background from primordial density
  %perturbations,''
  Phys.\ Rev.\  D {\bf 75}, 123518 (2007).

\bibitem{Bauetal}
  D.~Baumann, P.~J.~Steinhardt, K.~Takahashi and K.~Ichiki,
  %``Gravitational Wave Spectrum Induced by Primordial Scalar Perturbations,''
  Phys.\ Rev.\  D {\bf 76}, 084019 (2007).
  %%CITATION = PHRVA,D76,084019;%%


\bibitem{CooKamCal05}
  A.~Cooray, M.~Kamionkowski and R.~R.~Caldwell,
%``Cosmic shear of the microwave background: The curl diagnostic,''
  Phys.\ Rev.\ D {\bf 71}, 123527 (2005).
  %%CITATION = ASTRO-PH 0503002;%%


%\cite{Cooray:2002mj}
\bibitem{CooHu02}
A.~Cooray and W.~Hu,
%``Second Order Corrections to Weak Lensing by Large-Scale Structure,''
Astrophys.\ J.\  {\bf 574}, 19 (2002);
%%CITATION = ASTRO-PH 0202411;%%
  C.~M.~Hirata and U.~Seljak,
  %``Reconstruction of lensing from the cosmic microwave background
  %polarization,''
  Phys.\ Rev.\ D {\bf 68}, 083002 (2003).
  %%CITATION = ASTRO-PH 0306354;%%


\bibitem{WMAP3} 
%D.~N.~Spergel et al.,\ ApJS \ {\bf 170}, 377 (2007);
 E.~Komatsu {\it et al.},
  %``Five-Year Wilkinson Microwave Anisotropy Probe (WMAP) Observations:
  %Cosmological Interpretation,''
  arXiv:0803.0547 [astro-ph];
 J.~Dunkley {\it et al.}  [WMAP Collaboration],
  %``Five-Year Wilkinson Microwave Anisotropy Probe (WMAP) Observations:
  %Likelihoods and Parameters from the WMAP data,''
  arXiv:0803.0586 [astro-ph].
 
\bibitem{Prit}
  J.~R.~Pritchard and M.~Kamionkowski,
  %``Cosmic microwave background fluctuations from gravitational waves: An
  %analytic approach,''
  Annals Phys.\  {\bf 318}, 2 (2005).
  %%CITATION = ASTRO-PH 0412581;%%

\bibitem{CPT}
S.~M.~Carroll, W.~H.~Press and E.~L.~Turner,
  %``The Cosmological constant,''
  Ann.\ Rev.\ Astron.\ Astrophys.\  {\bf 30}, 499 (1992).
  %%CITATION = ARAAA,30,499;%%

\bibitem{spin2}
T.~Okamoto and W.~Hu,
  %``CMB Lensing Reconstruction on the Full Sky,''
  Phys.\ Rev.\ D {\bf 67}, 083002 (2003).
  

\end{thebibliography}
\end{document}